\documentclass[aps,pra,twocolumn,showpacs,superscriptaddress,floatfix]{revtex4-1}
\usepackage{amsmath}
\usepackage{amssymb,dsfont}
\usepackage[small]{subfigure}
\usepackage[dvips]{graphicx,color}
\usepackage{natbib}
\newcommand{\imu}{\textrm{i}}

\begin{document}
 \title{Dephasing effects on stimulated Raman adiabatic passage in tripod configurations}

\author{C. Lazarou} 
\affiliation{Department of Physics and Astronomy, University College London, Gower Street, London WC1E 6BT, United Kingdom}
\affiliation{Department of Physics, Sofia University, James Bourchier 5 blvd, 1164 Sofia, Bulgaria}
\author{N. V. Vitanov}
\affiliation{Department of Physics, Sofia University, James Bourchier 5 blvd, 1164 Sofia, Bulgaria}
\date{\today}

\begin{abstract}
We present an analytic description of the effects of dephasing processes on stimulated Raman 
adiabatic passage in a tripod quantum system. To this end, we develop an effective two-level model.
Our analysis makes use of the adiabatic approximation in the weak dephasing regime. An effective 
master equation for a two-level system formed by two dark states is derived, where analytic 
solutions are obtained by utilizing the Demkov-Kunike model. From these, it is found that the 
fidelity for the final coherent superposition state decreases exponentially for increasing dephasing rates. 
Depending on the pulse ordering and for adiabatic evolution the pulse delay can have an inverse effect. 
\end{abstract}
\maketitle
\section{Introduction} \label{sec1}
Stimulated Raman adiabatic passage (STIRAP) \cite{Bergmann1998,Vitanov2001,Vitanov2001b} is a powerful and robust technique
for achieving complete population transfer in three-state quantum systems. By using two pulsed laser fields population
is adiabatically transferred from an initially populated state $\psi_1$, to a target state $\psi_3$ 
via an intermediate state $\psi_2$. A unique feature of this technique is that the intermediate state is never 
populated. This is due to the fact that the system at all times adiabatically follows a dark state,
hence population losses due to spontaneous emission are suppressed.   

Apart from being used for population transfer, STIRAP can also be used to create coherent superpositions of states 
in quantum systems in 
tripod configurations \cite{Unanyan1998,Theuer1999,Vitanov2001}. The main idea behind this method is the same
as with STIRAP in $\Lambda$-configurations, with the exception that the system now adiabatically follows a 
superposition of two dark states. The interference of these two states results in a coherent superposition
between two or three of the ground states of the tripod system. The exact form for this  
final state is defined by the geometric phase acquired by the dark states.

In addition to the creation of coherent superpositions \cite{Unanyan1998,Theuer1999,Unanyan2004b} 
, STIRAP in tripod configurations can be further exploited to 
implement quantum gates \cite{Moller2007,Moller2008,Unanyan2004}. Furthermore, 
adiabatic passage in tripod systems can be used to engineer 
non-Abelian gauge potentials for ultracold atoms \cite{Ruseckas2005}. The formation
of such potentials is made possible because during the adiabatic evolution, the system 
can acquire non-Abelian phases \cite{Unanyan1999}. 

Since the intermediate state is not populated in the adiabatic limit, spontaneous emission from this state
is expected to have no effect on the fidelity. On the other hand, the creation of a coherent superposition 
relies on the formation of \textit{coherent} dark states. Thus maintaining coherence is vital for 
achieving higher fidelities. However phase relaxation effects induced for example by elastic collisions or 
laser phase fluctuations can have an adverse effect on the fidelity. 

Previous studies on STIRAP in the presence of dephasing \cite{Shi2003,Ivanov2004}, have shown that decoherence can 
lead to population losses from the dark state, resulting in a transfer efficiency reduction. On the other hand,
increasing the relative delay between the two laser pulses increases the transfer efficiency. 
This is due to the inverse dependence of the transition time with respect to the delay. In a recent paper
by \textit{M\o{}ller}, \textit{Madsen} and \textit{M\o{}lmer} \cite{Moller2008}, dephasing in tripod
systems and the effect it has on single qubit gates was considered. Using the Monte Carlo wavefunction
method, they were able to show that the system acquires complex geometric phases, implying losses 
which reduce the gate fidelity.     

In the present paper, we extend the method used in Ref. \cite{Ivanov2004} to study dephasing effects
on STIRAP in tripod configurations. The method makes use of the adiabatic approximation in the
weak dephasing regime, where we derive an effective two-level master equation for the dark states. Analytic 
solutions are obtained when the Stokes and control pulses overlap, whereas for other
pulse orderings we make use of numerical simulations. Depending on the pulse ordering, similar with or different features 
from STIRAP in a $\Lambda$-configuration are observed.

The paper is organized as follows. In Sec. \ref{sec2} we provide a brief introduction on STIRAP in tripod
configurations and the master equation is introduced. In Sec. \ref{sec3} we present the effective two-level model for
the dark states, and derive analytic solutions in Sec. \ref{sec4}. 
In Sec. \ref{sec5} we present results from numerical simulations. A summary of the results is given in Sec. \ref{sec5}.
\section{The tripod configuration} \label{sec2}
\begin{figure}[tbp]
  \begin{center}
    \resizebox{50mm}{!}{\includegraphics{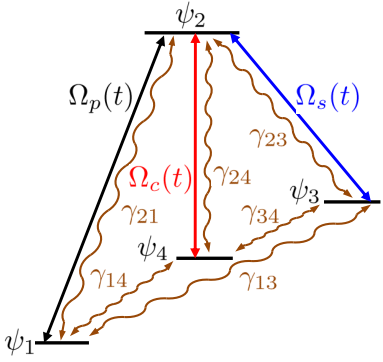}}
    \caption{(Color online) The tripod configuration and the three laser pulses: $\Omega_p(t)$ (pump), $\Omega_c(t)$ (control)
      and $\Omega_s(t)$ (Stokes). The relative dephasing for each pair of states is depicted by wavy lines.} \label{fig1}
  \end{center}
\end{figure}
The tripod system is shown in Fig. \ref{fig1}. The three ground states 
$\psi_1$, $\psi_3$ and $\psi_4$ are resonantly coupled to the intermediate level $\psi_2$ via
a pump $\Omega_p(t)$, a Stokes $\Omega_s(t)$ and a control pulse $\Omega_c(t)$ respectively. 
The system is initially prepared in state $\psi_1$, i.e.\ $\psi(-\infty)=\psi_1$.
Such a configuration was part of a proposed scheme for creating and phase probing 
coherent superpositions \cite{Unanyan1998}. This was demonstrated in an 
experiment by {\it Theuer} et al. \cite{Theuer1999}. 

The scheme we examine here uses an adiabatic passage method which is an extension of STIRAP in $\Lambda$-systems. 
The population initially placed in state $\psi_1$ is either 
transferred to the other two ground states $\psi_3$ and $\psi_4$ or is split between the states $\psi_1$
and $\psi_3$. The final superposition between $\psi_3$ and $\psi_4$, or between $\psi_3$ 
$\psi_1$, is determined by the relative distance of the pulses. The conditions for the robust creation
of such superpositions is that the system evolves adiabatically \cite{Unanyan1998,Vitanov2001}. 
\subsection{Adiabatic states: Creating coherent superpositions} \label{AdiabaticStates}
The interaction Hamiltonian in the rotating wave approximation reads \cite{Vitanov2001} 
\begin{equation} \label{eq1}
  H(t)=\frac{\hbar}{2}\left(\begin{array}{cccc}
    0 & \Omega_p(t) & 0 & 0 \\
    \Omega_p(t) & 0 & \Omega_s(t) & \Omega_c(t) \\
    0 & \Omega_s(t) & 0 & 0 \\
    0 & \Omega_c(t) & 0 & 0 
    \end{array}\right),
\end{equation}
where we take all transitions to be resonant with the respective laser field. 
Furthermore, we assume that these are the only allowed 
dipole transitions. This time-dependent Hamiltonian has two dark states $\Phi_1(t)$ and $\Phi_2(t)$ \cite{Vitanov2001}
\begin{subequations} \label{eq2}
  \begin{align}
    \nonumber
    \Phi_1(t)=&\frac{1}{\sqrt{2}}\psi_1\cos\theta-\frac{1}{\sqrt{2}}\psi_3(\sin\theta\cos\phi+\imu\sin\phi) \\
    &-\frac{1}{\sqrt{2}}\psi_4(\sin\theta\sin\phi-\imu\cos\phi), \label{eq2a} \\ \nonumber
    \Phi_2(t)=&\frac{1}{\sqrt{2}}\psi_1\cos\theta-\frac{1}{\sqrt{2}}\psi_3(\sin\theta\cos\phi-\imu\sin\phi)\\
    &-\frac{1}{\sqrt{2}}\psi_4(\sin\theta\sin\phi+\imu\cos\phi), \label{eq2b}
  \end{align}
\end{subequations}
where the time dependent mixing angles $\phi(t)$ and $\theta(t)$ read
\begin{subequations} \label{eq3}
  \begin{align}
    &\tan(\phi(t))=\frac{\Omega_c(t)}{\Omega_s(t)}, \label{eq3a} 
    \intertext{and}
    &\tan(\theta(t))=\frac{\Omega_p(t)}{\sqrt{\Omega^2_s(t)+\Omega^2_c(t)}}. \label{eq3b}
  \end{align}
\end{subequations}
We note here, that since the two dark states are eigenstates of $H(t)$ with zero eigenvalue, i.e.\
$H(t)\Phi_{1,2}(t)=0$, any linear superposition of these two states is also a dark state 
\cite{Vitanov2001,Unanyan1998}.
In addition to the two dark states, there are also two adiabatic states with non-zero time-dependent eigenenergies
\begin{subequations} \label{eq4}
  \begin{align}
    \nonumber
    \Phi_3(t)=&\frac{1}{\sqrt{2}}\psi_1\sin\theta+\frac{1}{\sqrt{2}}\psi_2+\frac{1}{\sqrt{2}}\psi_3\cos\theta\cos\phi \\
    &+\frac{1}{\sqrt{2}}\psi_4\cos\theta\sin\phi, \label{eq4a} \\ \nonumber 
    \Phi_4(t)=&\frac{1}{\sqrt{2}}\psi_1\sin\theta-\frac{1}{\sqrt{2}}\psi_2+\frac{1}{\sqrt{2}}\psi_3\cos\theta\cos\phi
    \\ &+\frac{1}{\sqrt{2}}\psi_4\cos\theta\sin\phi, \label{eq4b} 
  \end{align}
\end{subequations}
with eigenenergies
\begin{equation} \label{eq5}
  \epsilon_3(t)=-\epsilon_4(t)=\frac{\hbar}{2}\Omega(t),
\end{equation}
where $\Omega(t)=\sqrt{\Omega^2_p(t)+\Omega^2_s(t)+\Omega^2_c(t)}$ is the rms Rabi frequency.

In the adiabatic limit, the time-dependent eigenstates are weakly coupled 
and this is also valid for the two dark states $\Phi_1(t)$ and
$\Phi_2(t)$. Although they form a pair of degenerate states, $\epsilon_1(t)=\epsilon_2(t)=0$, the corresponding diabatic 
coupling is always zero, 
\begin{equation} \label{eq6}
  \langle\dot{\Phi}_1(t)\vert\Phi_2(t)\rangle=0,
\end{equation}
whereas
\begin{equation}
  \langle\Phi_{1}(t)\vert\dot{\Phi}_1(t)\rangle=-\langle\Phi_{2}(t)\vert\dot{\Phi}_2(t)\rangle=\imu\dot{\phi}\sin\theta.
\end{equation}
Because of this time-dependent energy shift, both dark states acquire a geometric phase $\vartheta_j$ \cite{Berry1984}
\begin{equation} \label{eq7}
  \vartheta_1=-\vartheta_2=\vartheta_g=\int_{-\infty}^{\infty}\textrm{d}t\dot{\phi}\sin\theta.
\end{equation}
Thus when the system starts in state $\Phi_1(t)$ or $\Phi_2(t)$, it will adiabatically follow this state and remain in this state 
at all times acquiring a net phase $\pm\vartheta_g$ respectively
\begin{equation} \label{eq8}
  \Phi_1(-\infty)\rightarrow e^{-\imu\vartheta_g}\Phi_1(\infty),\quad\Phi_2(-\infty)\rightarrow e^{\imu\vartheta_g}\Phi_2(\infty).
\end{equation}

As said in the begining of this section, the system is initially prepared in state $\psi_1$. Then in the adiabatic limit and for
$\theta(-\infty)=0$, the system state is a symmetric superposition of the two dark states
\begin{equation} \label{eq9}
  \Psi(-\infty)=\psi_1=\frac{1}{\sqrt{2}}\left(\Phi_1(-\infty)+\Phi_2(-\infty)\right),
\end{equation}
and for $t\rightarrow\infty$ it reads
\begin{equation} \label{eq10}
  \Psi(\infty)=\frac{1}{\sqrt{2}}\left(e^{-\imu\vartheta_g}\Phi_1(\infty)+e^{\imu\vartheta_g}\Phi_2(\infty)\right).
\end{equation}
The final superposition state in terms of the bare states depends on the ordering of the three pulses.
Of the many possible pulse orderings four different pulse sequences are particularly interesting for 
each produces a different coherent state \cite{Vitanov2001}. These pulse orderings are as follows.
\begin{itemize}
  \item The pulses are ordered so that the Stokes pulse starts first and ends after the pump pulse, while the
    control pulse is delayed with respect to both of them. For this pulse sequence we have the following 
    asymptotic relations: $\theta(-\infty)=\theta(\infty)=0$, $\phi(-\infty)=0$ and $\phi(\infty)=\pi/2$.
    Then the final state reads
    \begin{equation} \label{eq11}
      \Psi(\infty)=\psi_1\cos\vartheta_g-\psi_3\sin\vartheta_g.
    \end{equation}
  \item A different pulse ordering is arranged so that the Stokes pulse comes first, followed by the 
    control pulse and then the pump pulse. For this case the asymptotic relations are $\theta(-\infty)=0$,
    $\theta(\infty)=\pi/2$, $\phi(-\infty)=0$ and $\phi(\infty)=\pi/2$. Then the final state is
    \begin{equation} \label{eq12}
      \Psi(\infty)=-\psi_3\sin\vartheta_g-\psi_4\cos\vartheta_g.
    \end{equation}
  \item Alternatively one can reverse the order of the Stokes and control pulses, i.e.\ the latter pulse 
    precedes the Stokes pulse. 
    Then for $t\rightarrow\pm\infty$ we have
    $\theta(-\infty)=0$, $\theta(\infty)=\pi/2$, $\phi(-\infty)=\pi/2$ and $\phi(\infty)=0$. The final
    coherent state reads
    \begin{equation} \label{eq13}
      \Psi(\infty)=-\psi_3\cos\vartheta_g+\psi_4\sin\vartheta_g.
    \end{equation}
  \item Finally, the Stokes and control pulses can coincide in time and precede the pump pulse. Then the
    following asymptotic relations apply: $\theta(-\infty)=0$, $\theta(\infty)=\pi/2$ and 
    $\phi(-\infty)=\phi(\infty)=\pi/4$. Because $\phi(t)$ is constant, we have that $\dot{\phi}(t)=0$ and thus
    the geometric phase is zero. Hence the final state is 
    \begin{equation} \label{eq14}
      \Psi(\infty)=-\frac{1}{\sqrt{2}}(\psi_3+\psi_4).
    \end{equation}
\end{itemize}
\subsection{Dephasing} \label{Dephasing}
In order to model the effect of dephasing in the tripod system, we make use of the master (Liouville) equation
\begin{equation} \label{eq15}
  \imu\hbar\dot{\rho}=[H(t),\rho]+D.
\end{equation}
The dissipator matrix $D$ describes dephasing effects
\begin{equation} \label{eq16}
  D=-\imu\hbar\left(\begin{array}{cccc}
    0 & \gamma_{12}\rho_{12} & \gamma_{13}\rho_{13} & \gamma_{14}\rho_{14} \\
    \gamma_{21}\rho_{21} & 0 & \gamma_{23}\rho_{23} & \gamma_{24}\rho_{24} \\
    \gamma_{31}\rho_{31} & \gamma_{32}\rho_{32} & 0 & \gamma_{34}\rho_{34} \\
    \gamma_{41}\rho_{41} & \gamma_{42}\rho_{42} & \gamma_{43}\rho_{43} & 0
    \end{array}\right),
\end{equation}
where $\gamma_{ij}=\gamma_{ji}$ are the constant relaxation rates and $\rho$ is the density 
matrix in the bare basis $\psi_m$, i.e.\ $\rho_{mn}=\langle\psi_m\vert\hat{\rho}\vert\psi_n\rangle$.
The initial conditions are $\rho_{11}(-\infty)=1$ and $\rho_{mn}=0$ for $mn\neq11$.

The derivation of master equations such as the one in Eq. (\ref{eq15}) is based on the
use of the Born-Markov approximation \cite{Breuer}. This imposes restrictions on
the spectral properties of the heat bath with which a quantum system interacts \cite{Lambropoulos2000}, while 
the relative coupling strength must be weak. Thus, collisions 
between atoms or molecules must be weak \cite{Stenholm}, whereas the fluctuating laser 
phase must be well approximated by a Markovian process \cite{Agarwal1978}.

In the following section we derive approximate solutions in the weak dephasing
limit and for adiabatic evolution. Because of this latter assumption we will be using the density matrix
in the adiabatic basis, defined from the following transformation
\begin{equation} \label{eq17}
  \rho^a=R^{-1}\rho R
\end{equation}
where $\rho^a$ is the density matrix in the adiabatic basis 
$\Phi_j(t)$, i.e.\ $\rho^a_{mn}=\langle\Phi_m(t)\vert\hat{\rho}\vert\Phi_n(t)\rangle$. The rotation matrix $R$ 
is formed by using the adiabatic states as its columns
\begin{equation} \label{eq18}
  R(t)=\left[\Phi^T_1(t),\Phi^T_2(t),\Phi^T_3(t),\Phi^T_4(t)\right],
\end{equation}
where $R^{-1}=R^\dagger$. Then the master equation in the adiabatic basis is
\begin{equation}\label{eq19}
  \imu\hbar\dot{\rho}^a=[H^a(t),\rho^a]-\imu\hbar[R^{-1}\dot{R},\rho^a]+R^{-1}DR,
\end{equation}
where $H^a(t)$ is the Hamiltonian in the adiabatic basis,
\begin{equation} \label{eq20}
  H^a(t)=\textrm{diag}(0,0,\epsilon_3(t),\epsilon_4(t)).
\end{equation}
The second term on the right hand side of Eq. (\ref{eq19}), i.e.\ $R^{-1}\dot{R}$, 
describes non-adiabatic interactions (off diagonal terms), whereas the two diagonal terms
$(R^{-1}\dot{R})_{11}=-(R^{-1}\dot{R})_{22}$ are responsible for the geometric phases acquired
by the two dark states.
\section{The two-level approximation} \label{sec3}
As mentioned earlier, the creation of coherent states is
the result of interference effects between the two dark states (\ref{eq2}). This suggests
that it is possible to solve the master equation (\ref{eq15}) approximately by deriving an effective two-level
master equation. To this end, we make two approximations: the adiabatic 
and the weak dephasing approximations.
These were used before when studying the effect of dephasing on STIRAP in 
$\Lambda$-systems \cite{Ivanov2004}.
\subsection{The adiabatic approximation} \label{sec31}
The first of the two approximations that we are using is that of adiabatic evolution. In order for
this to be valid we need large pulse areas \cite{Unanyan1998,Vitanov2001} so that
\begin{equation} \label{eq21}
  \vert\dot{\theta}(t)\vert\ll\vert\Omega(t)\vert,\qquad\vert\dot{\phi}(t)\vert\ll\vert\Omega(t)\vert.
\end{equation}
When these two conditions are satisfied, the system adiabatically follows a superposition of the two dark states,
whereas the adiabatic states $\Phi_3(t)$ and $\Phi_4(t)$ are not populated. Because of this
the coherences $\rho^a_{ij}$ related to these two states are negligible.
\subsection{The weak dephasing approximation} \label{sec32}
The second approximation is that of weak dephasing, i.e.\ the relaxation rates $\gamma_{ij}$ are 
much smaller than the rms Rabi frequency $\Omega(t)$
\begin{equation} \label{eq22}
  \gamma_{ij}\ll\vert\Omega(t)\vert,\qquad (i,j=1,2,3,4;i\neq j).
\end{equation}
Assuming that the adiabatic approximation Eq. (\ref{eq21}) is valid, then for $\gamma_{ij}\ge\vert\Omega(t)\vert$ we
would have that $\gamma_{ij}T\gg1$, with $T$ a characteristic time length for the pulse durations. 
This latter inequality corresponds to strong dephasing which would lead to complete incoherent dynamics that
are governed by rate equations \cite{Shore}. This justifies the use of the weak dephasing approximation 
Eq. (\ref{eq22}).

Using both approximations we can now simplify the analysis by neglecting all coherences that include the
two adiabatic states $\Phi_3(t)$ and $\Phi_4(t)$,
\begin{equation} \label{eq23}
  \rho^a_{ij}\approx0,\qquad (ij\neq12,21).
\end{equation}
With this the density matrix in the adiabatic basis $\rho^a$ acquires the following approximate form
\begin{equation} \label{eq24}
  \rho^a\approx\left(\begin{array}{cccc}
    \rho^a_{11} & \rho^a_{12} & 0 & 0 \\
    \rho^a_{21} & \rho^a_{22} & 0 & 0 \\
    0 & 0 & \rho^a_{33} & 0 \\
    0 & 0 & 0 & \rho^a_{44} 
    \end{array}\right).
\end{equation}
Using Eqs. (\ref{eq17}), (\ref{eq18}), (\ref{eq19}) and (\ref{eq24}) (see appendix \ref{appA}),
we can first show that the population inversion for the bright states, $w^a_{34}(t)=\rho^a_{33}-\rho^a_{44}$,
and for the dark states, $w^a_{12}(t)=\rho^a_{11}-\rho^a_{22}$, are both zero at all times, i.e.\
\begin{equation} \label{eq25}
\rho^a_{11}=\rho^a_{22},\qquad\rho^a_{33}=\rho^a_{44}.
\end{equation}
Hence, the populations for the adiabatic states can be parametrized in terms of a single function $s(t)$, i.e.\
\begin{subequations} \label{eq26}
  \begin{align}
    \rho^a_{11}&=\rho^a_{22}=\frac{1-2s(t)}{2}, \label{eq26a}
    \\ \nonumber \\
    \rho^{a}_{33}&=\rho^a_{44}=\frac{1+2s(t)}{2}. \label{eq26b}
  \end{align}
\end{subequations}
From this we can derive a set of coupled differential equations for $s(t)$ and the coherences for the dark states, 
$u(t)=\sqrt{2}\textrm{Re}\{\rho^a_{12}\}$ and $v(t)=\sqrt{2}\textrm{Im}\{\rho^a_{12}\}$,
\begin{subequations}\label{eq27}
  \begin{align}
    \dot{s}(t)=&-\Gamma_{s}(t)s(t)+\sqrt{2}\Omega_{su}(t)u(t)+\sqrt{2}\Omega_{sv}(t)v(t), \label{eq27a}\\
    \nonumber
    \dot{u}(t)=&-\Gamma_{u}(t)u(t)+\left(2\dot{\phi}\sin\theta+\Omega_{uv}(t)\right)v(t)
    \\ &+\sqrt{2}\Omega_{su}(t)s(t), \label{eq27b} \\
    \nonumber
    \dot{v}(t)=&-\Gamma_{v}(t)v(t)+\left(-2\dot{\phi}\sin\theta+\Omega_{uv}(t)\right)u(t)
    \\ &+\sqrt{2}\Omega_{sv}(t)s(t). \label{eq27c}
  \end{align}
\end{subequations}   
The initial conditions are $s(t)=-1/2$, $u(t)=1/\sqrt{2}$ and $v(t)=0$.
The derivation of these equations and the expressions for the Rabi frequencies $\Omega_{su}(t)$, $\Omega_{sv}(t)$, $\Omega_{uv}(t)$, 
and those for the relaxation rates $\Gamma_j(t)$, $j=s,u,v$, are provided in appendix \ref{appA}.
We note here that the effective relaxation rates $\Gamma_j(t)$ and the Rabi frequencies $\Omega_{ij}(t)$, depend only
on the relaxation rates $\gamma_{13}$, $\gamma_{14}$ and $\gamma_{34}$, see Eqs. (\ref{eqA8}) and (\ref{eqA9}).
\subsection{Fidelity for the target coherent superposition state} \label{sec33}
In the analysis that follows we will be using the fidelity $F(t)$ for the final coherent state 
$\Psi(\infty)$ \cite{Nielsen}
\begin{equation} \label{eq28}
  F^2(t)=\vert\langle\Psi(\infty)\vert\rho(t)\vert\Psi(\infty)\rangle\vert^2,
\end{equation}
where $\Psi(\infty)$ is one of the four states (\ref{eq11}), (\ref{eq12}), (\ref{eq13}) or (\ref{eq14}), and
$\rho(t)$ is the density matrix in the bare basis. For $t\rightarrow\infty$ the fidelity can be expressed in terms
of the dark state populations and coherences.
Using Eq. (\ref{eq10}) for the coherent state $\Psi(\infty)$, and Eq. (\ref{eq24}) for the density matrix 
in the adiabatic basis and for weak dephasing, the fidelity reads
\begin{equation} \label{eq29}
  \begin{split}
    F^2(\infty)=&\rho^a_{11}(\infty)+\cos(2\vartheta_g)\textrm{Re}\{\rho^a_{12}(\infty)\} \\ ~\\
    & -\sin(2\vartheta_g)\textrm{Im}\{\rho^a_{12}(\infty)\},
  \end{split}
\end{equation}
where $\vartheta_g$ is given by Eq. (\ref{eq7}).

To this end, and before proceeding with the solution of Eqs. (\ref{eq27}), we introduce one more tool that we will
be using in the following sections. This is the transition time $T_{tr}(\epsilon)$, and is defined as the time 
needed for the fidelity to rise from a small value $F^2(t_\epsilon)=\epsilon$ to $F^2(t_{1-\epsilon})=1-\epsilon$, i.e.\ 
\begin{equation} \label{eq30}
  T_{tr}(\epsilon)=t_{1-\epsilon}-t_{\epsilon}.
\end{equation}
The importance of the transition time was discussed on a previous work on dephasing effects on STIRAP in $\Lambda$-systems 
\cite{Ivanov2004}. For STIRAP the pulse delay has an inverse effect on the efficiency, and this is
due to the fact that the transition time is inversely proportional to the delay time.

We should note that the above definition for the transition time, does not apply when the Stokes pulse 
ends after the pump pulse, see Eq. (\ref{eq11}). While for the other three possible pulse orderings the fidelity 
is initially zero, i.e.\ $F^2(-\infty)=0$, for the former pulse ordering the fidelity is $F^2(-\infty)=\cos^2(\vartheta_g)$. 
For this case, instead of using the above definition for the transition time, we will be using the following one
\begin{equation} \label{eq31}
  T_{tr}(\epsilon)=t_{1-\epsilon}-t_{1+\epsilon},
\end{equation}
where the time $t_{1+\epsilon}$ is such that 
\begin{equation}
  F^2(t_{1+\epsilon})=(1+\epsilon)F^2(-\infty), 
\end{equation}
and $t_{1-\epsilon}$ remains the same.
\section{Analytic solutions} \label{sec4}
\subsection{Equal relaxation rates $\gamma_{ij}=\gamma$ and $\Omega_s(t)=\Omega_c(t)$} \label{sec41}
\begin{figure}[tbp]
  \begin{center}
    \resizebox{85mm}{!}{\includegraphics{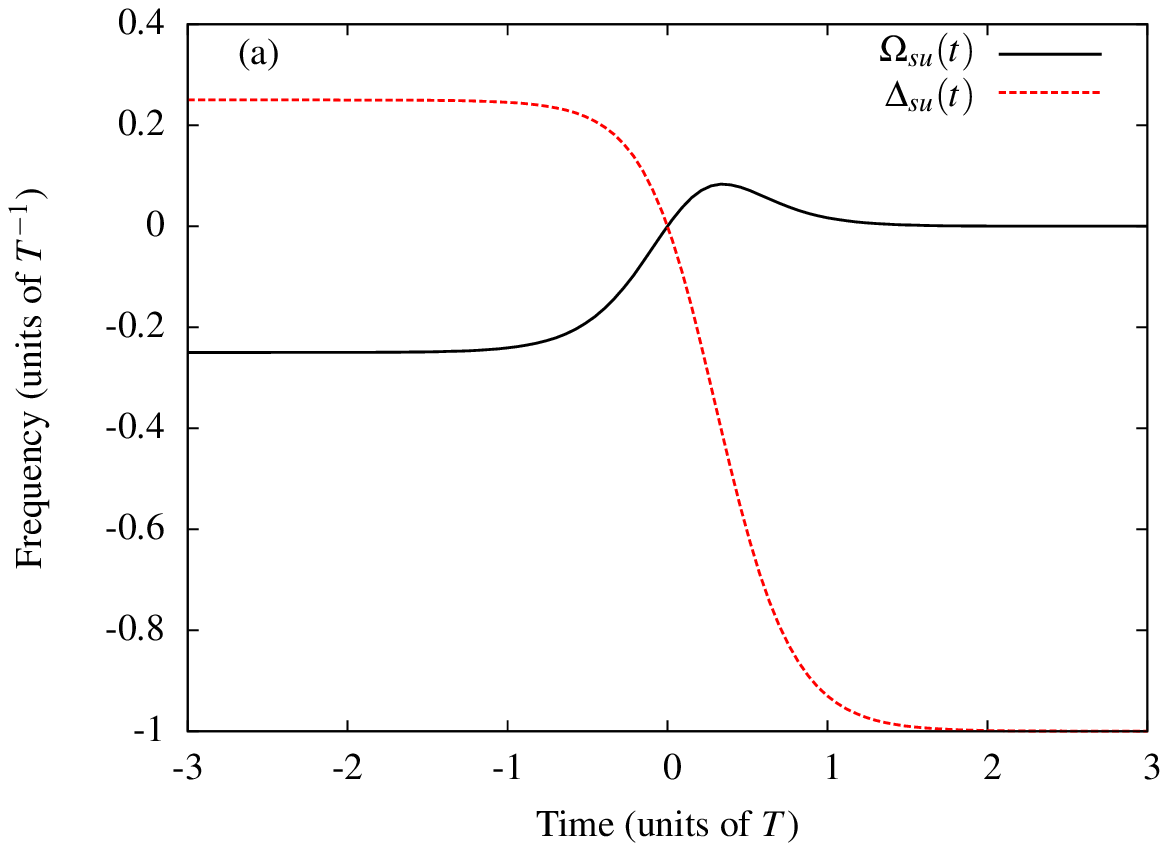}} \\~\\
    \resizebox{85mm}{!}{\includegraphics{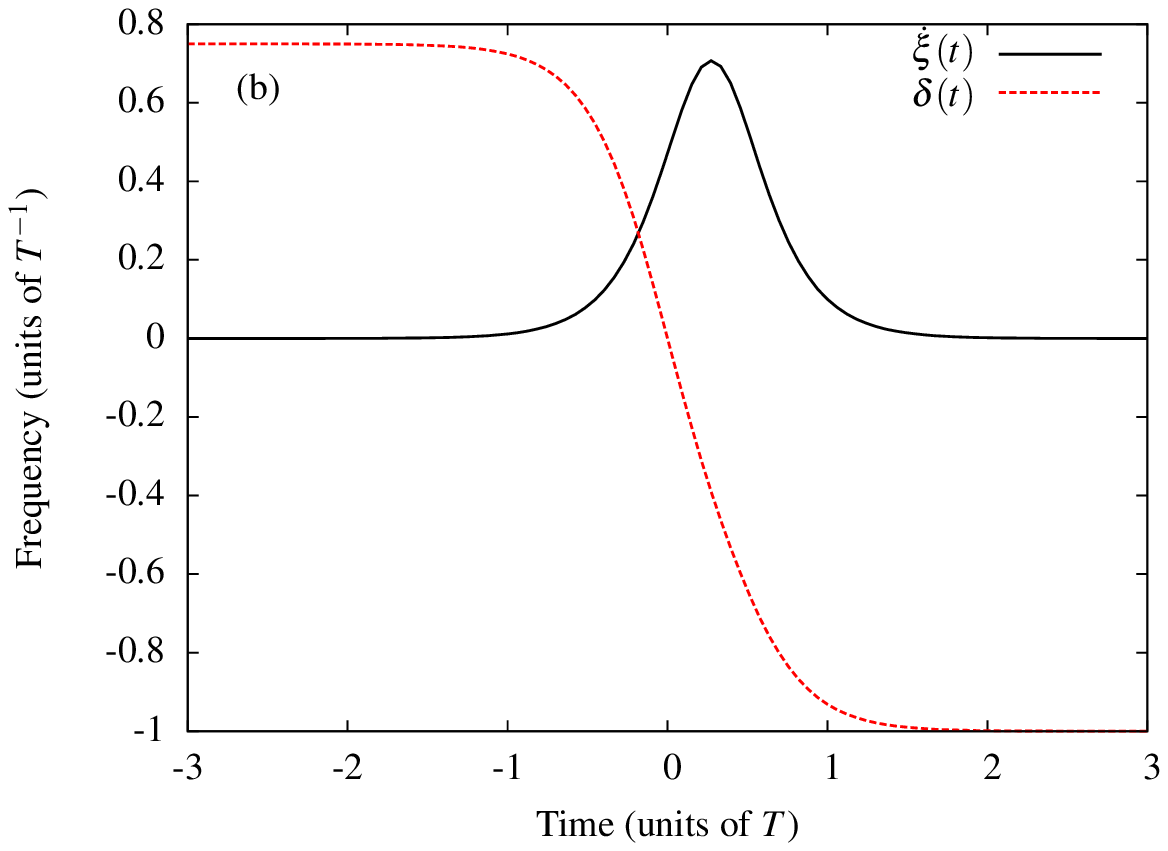}}
    \caption{(Color online) (a) The coupling $\Omega_{su}(t)$ and the detuning $\Delta_{su}(t)$ Eq. (\ref{eq33}) for 
      Gaussian pulses Eq. (\ref{eq42}). (b) The coupling $\dot{\xi}(t)$ and the detuning $\Delta(t)$ Eq. (\ref{eq43}) 
      in the adiabatic basis (\ref{eq41}). The delay is $\tau=T$ and $\gamma=T^{-1}$.}\label{fig2}
  \end{center}
\end{figure}

A special case is that when the Stokes and control pulse coincide, i.e.\ $\Omega_c(t)=\Omega_s(t)$. 
Then at all times we have $\phi(t)=\pi/4$ and $\dot{\phi}(t)=0$. When taking all relaxation
rates equal i.e.\ $\gamma_{ij}=\gamma$ we have that
\begin{equation} \label{eq32}
  \Omega_{sv}(t)=\Omega_{uv}(t)=0,
\end{equation}
and the equations for $s(t)$ and $u(t)$ decouple from that for $v(t)$, i.e.\ 
\begin{subequations}\label{eq33}
  \begin{align}
    &\dot{c}_1(t)=\sqrt{2}\Omega_{su}(t)c_2(t), \label{eq33a}\\
    &\dot{c}_2(t)=\Delta_{su}(t)c_2(t)+\sqrt{2}\Omega_{su}(t)c_{1}(t), 
    \label{eq33b} \intertext{and for $v(t)$ we have}
    &\dot{v}(t)=-\Gamma_{v}(t)v(t). \label{eq33c}
  \end{align}
\end{subequations}
The new variables $c_1(t)$ and $c_2(t)$ are 
\begin{equation} \label{eq34}
  \begin{split}
    c_1(t)&=s(t)\exp\left(\int^t\Gamma_s(t')\textrm{d}t'\right), \\
    c_2(t)&=u(t)\exp\left(\int^t\Gamma_s(t')\textrm{d}t'\right).
  \end{split}
\end{equation}
This parametrization is used in order to emphasize the analogy to a two-level system.
The effective relaxation rates and coupling in terms of the mixing angle $\theta(t)$ read
\begin{subequations} \label{eq35}
  \begin{align}
    &\Gamma_v(t)=\gamma\cos^2(\theta), \label{eq35a} \\
    &\Omega_{su}(t)=
    \frac{\gamma}{4}\left(\frac{3\sin^2(2\theta)}{4}-\cos^2\theta\right),\label{eq35c} \\
    &\Delta_{su}(t)=
    \frac{\gamma}{4}\left(\frac{3\sin^2(2\theta)}{4}+\cos^2\theta\right)-\gamma\sin^2\theta. \label{eq35b} 
  \end{align}
\end{subequations}
In Fig. \ref{fig2}(a), we plot the coupling $\Omega_{su}(t)$ and the detuning $\Delta_{su}(t)$ for
Gaussian pulses Eq. (\ref{eq42}).
To this end, we note that since $v(-\infty)=0$ equation (\ref{eq33c}) has the trivial solution $v(t)=0$.

In order to connect Eqs. (\ref{eq33a}) and (\ref{eq33b}) with those for a driven two-level system
and exploit this to solve them, we make use of the following time-dependent rotation 
\begin{equation} \label{eq36}
  \mathcal{R}(t)=\frac{2}{\sqrt{3}}\left(\begin{array}{cc}
    -\cos\xi & \sin\xi \\ 
    \sin\xi & \cos\xi
    \end{array}\right),
\end{equation}
where the angle $\xi(t)$ is
\begin{equation} \label{eq37}
  \xi(t)=\frac{1}{2}\arctan\left(\frac{2\sqrt{2}\Omega_{su}(t)}{\Delta_{su}(t)}\right).
\end{equation}
With this equations (\ref{eq33a}) and (\ref{eq33b}) take the form
\begin{equation} \label{eq38}
  \left(\begin{array}{c}
    \dot{c}_-(t) \\ \dot{c}_+(t) \end{array}\right)
  =\left(\begin{array}{cc}
    \epsilon_-(t) & \dot{\xi}(t) \\
    -\dot{\xi}(t) & \epsilon_+(t)
  \end{array}\right)\left(\begin{array}{c}
    c_-(t) \\ c_+(t) \end{array}\right),
\end{equation} 
where the ``energies''
\begin{equation} \label{eq39}
  \epsilon_{\pm}(t)=\frac{\Delta_{su}(t)}{2}\left(1\pm\sec(2\xi)\right),
\end{equation}
are the eigenvalues of the matrix 
\begin{equation} \label{eq40}
  W(t)=\left(\begin{array}{cc}
    0 & \sqrt{2}\Omega_{su}(t) \\
    \sqrt{2}\Omega_{su}(t) & \Delta_{su}(t)\end{array}\right).
\end{equation}
The corresponding eigenstates $\psi_{\pm}(t)$ are
\begin{equation} \label{eq41}
  \begin{split}
    \psi_{+}(t)&=\left[\sin\xi,\cos\xi\right]^T, \\
    \psi_{-}(t)&=\left[-\cos\xi,\sin\xi\right]^T.
  \end{split}
\end{equation}
Multiplying now both sides of Eq. (\ref{eq38}) with the imaginary unit, we see that 
it takes the form of a Schr\"odinger equation for a two-level system. The two states 
$\psi_-$ and $\psi_+$, are driven by a laser field with Rabi frequency $\dot{\xi}(t)$, 
while decaying with rates $\epsilon_-(t)$ and $\epsilon_+(t)$ respectively. 
The initial conditions for Eq. (\ref{eq38}) can be derived from 
Eqs. (\ref{eq36}), (\ref{eq37}), (\ref{eq35b}) and (\ref{eq35c}), and are 
$c_+(-\infty)=1$ and $c_-(-\infty)=0$. 

In Fig. \ref{fig2}(b) we plot the ``chirped'' detuning $\Delta(t)=\epsilon_+(t)-\epsilon_-(t)$ and the coupling $\dot{\xi}(t)$ for 
Gaussian pulses of the form
\begin{equation} \label{eq42}
  \begin{split}
    &\Omega_p(t)=\Omega_0e^{-(t-\tau/2)^2/T^2}, \\
    &\Omega_s(t)=\Omega_c(t)=\Omega_0e^{-(t+\tau/2)^2/T^2}.
  \end{split}
\end{equation}
The ``chirped'' detuning $\Delta(t)$ and the coupling $\dot{\xi}(t)$ are
\begin{equation} \label{eq43}
  \begin{split}
    \Delta(t)&=\gamma f_1\left(\frac{4t\tau}{T^2}\right), \\ \\
    \dot{\xi}(t)&=\frac{\tau}{T^2} f_2\left(\frac{4t\tau}{T^2}\right),
  \end{split}
\end{equation}
where
\begin{equation} \label{eq44}
  \begin{split}
    f_1(x)&=\frac{1-e^{2x}}{(2+e^x)^2}\sqrt{1+\frac{8}{(1+e^x)^2}}, \\ \\
    f_2(x)&=\frac{2\sqrt{2}}{1+5\cosh(x)-4\sinh(x)}.
  \end{split}
\end{equation}
From Fig. \ref{fig2}, we see that both $\dot{\xi}(t)$ and $\Delta(t)$
resemblance the laser field and the frequency chirp for 
the Demkov-Kunike (DK) model \cite{Demkov1969}. This feature is 
exploited next to derive analytic solutions for Eqs. (\ref{eq38}). 
To this end, it should be pointed out that the time-dependent detuning in
the original DK model corresponds to real frequency chirp, whereas in the
two-level system of Eq. (\ref{eq38}) it is an imaginary chirp, i.e.\
a time-dependent decay rate.
\subsubsection{Adiabatic following} \label{sec411}
The method we use to solve Eq. (\ref{eq38}), is the one used in Refs. \cite{Rangelov2005,Ivanov2008}
to derive the propagator in three-state systems with pairwise crossings. Starting with 
the system initially in state $\psi_+(\infty)$, we assume that it evolves adiabatically 
until reaching the crossing point $t=0$, where $\Delta(0)=0$. At this point, diabatic transitions
will occur and the new system state will be a mixture of the two adiabatic states $\psi_\pm(t)$. The 
effect of the crossing is expressed in terms of a transition matrix
\begin{equation} \label{eq45}
  U_c(0)=\left(\begin{array}{cc}
    U_{--}(0) & U_{-+}(0) \\
    U_{+-}(0) & U_{++}(0)
    \end{array}\right),
\end{equation}
where $U_{jj}(0)$ and $U_{ij}(0)$ ($j\neq i$) are the survival and transition probability amplitudes, 
for a two-level system driven by a laser pulse $\dot{\xi}_{eff}(t)\approx\dot{\xi}(t)$, 
in the presence of a time-dependent spontaneous emission $\Delta_{eff}(t)\approx\Delta(t)$.

For times $t>0$, i.e.\ after the crossing the system will evolve adiabatically. With this the final
system propagator $U(\infty,-\infty)$ reads
\begin{equation} \label{eq46}
  U(\infty,-\infty)=U_a(\infty,0)U_c(0)U_a(0,-\infty),
\end{equation}
where the adiabatic propagator $U_a(t_f,t_i)$ is
\begin{equation} \label{eq47}
  U_a(t_f,t_i)=\left(\begin{array}{cc}
    e^{\int_{t_i}^{t_f}\epsilon_-(t)dt} & 0 \\
    0 & e^{\int_{t_i}^{t_f}\epsilon_+(t)dt} \end{array}\right).
\end{equation}
Upon using the above equations with the initial condition $\psi(-\infty)=\psi_+(-\infty)$, the
final two-level state takes the form
\begin{equation} \label{eq48}
  \begin{split}
    \psi(\infty)=&U_{-+}(0)e^{\int_{0}^{\infty}\epsilon_-(t)dt+\int_{-\infty}^0\epsilon_+(t)dt}\psi_-(\infty) \\ ~\\
    &+U_{++}(0)e^{\int_{-\infty}^{\infty}\epsilon_+(t)dt}\psi_+(\infty).
  \end{split}
\end{equation}
\subsubsection{Diabatic transitions} \label{sec412}
The survival probability amplitude $U_{++}(0)$ and the transition amplitude $U_{-+}(0)$, are solutions 
of the Schr\"odinger equation
\begin{equation} \label{eq49}
  \begin{split}
    &\dot{c}_-(t)=\dot{\xi}_{eff}(t)e^{\int^t\Delta_{eff}(t')dt'}c_+(t), \\ ~\\
    &\dot{c}_+(t)=-\dot{\xi}_{eff}(t)e^{-\int^t\Delta_{eff}(t')dt'}c_-(t).
\end{split}
\end{equation}
As already pointed out earlier and shown in Fig. \ref{fig2}(b),
this two-level system resemblance the Demkov-Kunike model \cite{Demkov1969}, with the only difference 
that the real frequency chirp for the latter model is replaced by an imaginary one, i.e.\ population loss.

To proceed with the solution of Eq. (\ref{eq49}) with the initial condition $c_+(-\infty)=1$ and $c_-(-\infty)=0$,
we utilize a hyperbolic sech function to approximate $\dot{\xi}(t)$
\begin{subequations} \label{eq50}
  \begin{align}
    &\dot{\xi}_{eff}(t)=A\textrm{sech}\left((t-t_{max})/T_{eff}\right), \label{eq50a} 
    \intertext{and a hyperbolic $\tanh$ function with a constant term for $\Delta(t)$}
    &\Delta_{eff}(t)=D+B\tanh\left((t-t_{max})/T_{eff}\right). \label{eq50b}
  \end{align}
\end{subequations}
The time $t_{max}$ refers to the maximum of $\dot{\xi}(t)$ Eq. (\ref{eq43}) and is
\begin{equation} \label{eq51}
  t_{max}=\frac{T^2}{4\tau}\textrm{arctanh}\left(\frac{4}{5}\right).
\end{equation}

The amplitude $A$ for $\dot{\xi}_{eff}(t)$ is derived from the condition $\dot{\xi}(t_{max})=\dot{\xi}_{eff}(t_{max})$,
and is 
\begin{equation} \label{eq52}
  A=\frac{\tau}{\sqrt{2}T^2}.
\end{equation}
The effective pulse duration $T_{eff}$ is obtained by requiring that both $\dot{\xi}(t)$ and $\dot{\xi}_{eff}(t)$
have the same pulse area. With this $T_{eff}$ reads
\begin{equation} \label{eq53}
  T_{eff}=\frac{T^2}{\sqrt{2}\pi\tau}\textrm{arctan}(2\sqrt{2}).
\end{equation}
Finally, the condition for obtaining $D$ and $B$, is that for $t\approx t_{max}$, $\Delta(t)\approx\Delta_{eff}(t)$. After 
performing a Taylor series expansion for both $\Delta(t)$ and $\Delta_{eff}(t)$ at the vicinity of $t_{max}$ and 
keeping only terms of first order in $t$, we have
\begin{subequations} \label{eq54}
  \begin{align}
    &D=-\frac{4\sqrt{6}\gamma}{25}, \label{eq54a}
    \intertext{and}
    &B=-\frac{64\sqrt{3}\gamma}{125\pi}\textrm{arctan}(2\sqrt{2}). \label{eq54b}
  \end{align}
\end{subequations}

Solutions for Eq. (\ref{eq49}) and for the Rabi frequency $\dot{\xi}_{eff}(t)$ Eq. (\ref{eq50a}) 
and for $\Delta_{eff}(t)$ Eq. (\ref{eq50b}), can be expressed in terms of hypergeometric functions, 
see for example Refs. \cite{Suominen1992,Vitanov1998}. For $t\rightarrow\infty$ the probability amplitudes 
are
\begin{subequations} \label{eq55}
  \begin{align}
    \nonumber
    U_{++}(0)=&\frac{\Gamma\left(\frac{1}{2}+\delta-\beta\right)}
    {\Gamma\left(\frac{1}{2}+\delta+\sqrt{\beta^2+\alpha^2}\right)}
    \\ &\times
    \frac{\Gamma\left(\frac{1}{2}+\delta+\beta\right)}
    {\Gamma\left(\frac{1}{2}+\delta-\sqrt{\beta^2+\alpha^2}\right)}, 
    \label{eq55a} \\ \nonumber \\
    \nonumber
    U_{-+}(0)=&
    \frac{\alpha\Gamma\left(\frac{1}{2}+\delta-\beta\right)}{\Gamma\left(1-\beta+\sqrt{\beta^2+\alpha^2}\right)}
    \\ &\times
    \frac{\Gamma\left(\frac{1}{2}-\delta-\beta\right)}{\Gamma\left(1-\beta-\sqrt{\beta^2+\alpha^2}\right)},
    \label{eq55b}
  \end{align}
\end{subequations}
where $\Gamma(x)$ is the gamma function \cite{Gradshteyn}. The parameters $\alpha$, $\beta$ and $\delta$ are
\begin{subequations} \label{eq56}
  \begin{align}
    &\alpha=AT_{eff}=\frac{1}{2\pi}\textrm{arctan}(2\sqrt{2}),\label{eq56a} \\ \nonumber \\
    &\beta=\frac{BT_{eff}}{2}=-\frac{16\sqrt{6}\textrm{arctan}^2(2\sqrt{2})}{125\pi^2}\frac{\gamma T^2}{\tau}, \label{eq56b} \\ 
    \nonumber \\
    &\delta=\frac{DT_{eff}}{2}=-\frac{2\sqrt{3}\textrm{arctan}(2\sqrt{2})}{25\pi}\frac{\gamma T^2}{\tau}. \label{eq56c}
  \end{align}
\end{subequations}
\subsubsection{Populations and coherences for the dark states} \label{sec413}
After substituting Eqs. (\ref{eq55a}) and (\ref{eq55b}) into Eq. (\ref{eq48}), and using the inverse rotation 
$\mathcal{R}^{-1}(\infty)$ along with Eqs. (\ref{eq34}), we obtain $s(\infty)$ and $u(\infty)$. From these we 
derive the real part for the coherence $\rho^a_{12}(\infty)$ of the two-dark states (\ref{eq2}), 
\begin{subequations} \label{eq57}
  \begin{align}
    &\textrm{Re}\{\rho^a_{12}(\infty)\}=\lim_{t\rightarrow\infty}\sqrt{\frac{3}{8}}U_{++}(0)
    \exp\left(-\frac{c_u\gamma T^2}{4\tau}-\gamma t\right), \label{eq57a}
    \intertext{and their populations}
    &\rho^{a}_{11}(\infty)=\rho^{a}_{22}(\infty)=\frac{1}{4}+\frac{\sqrt{3}}{4}U_{-+}(0)
         \exp\left(-\frac{c_s\gamma T^2}{4\tau}\right), \label{eq57b}
  \end{align}
\end{subequations}
where $c_s=2.42$ and $c_u=0.68$. The exponential terms in the above expressions are contributions acquired during the adiabatic
evolution, see Eq. (\ref{eq48}). Their dependence on the different parameters, i.e.\ the factor $\gamma T^2/\tau$ can be 
easily derived, whereas the two factors $c_u$ and $c_s$ can be calculated numerically, see appendix \ref{appB}.

We should recall here that the imaginary part of $\rho^a_{12}(\infty)$ is zero, because $v(t)=0$. 
Furthermore we note that in the long time limit the real part is also zero. Then, the fidelity for a coherent 
state (\ref{eq29}),  will be proportional to the population of the dark states. As we can see from Eq. (\ref{eq57b})
this will decrease for increasing relaxation rate $\gamma$, or will increase for increasing delay $\tau$. This is because of the 
inverse dependence of the transition time with respect to $\tau$. The transition time $T_{tr}(\epsilon)$ for $\epsilon=0.1$
can be derived from the following expression for the fidelity
\begin{equation}
  F^2(t)=\sin^2\theta=\frac{\exp\left(\frac{4t\tau}{T^2}\right)}{2+\exp\left(\frac{4t\tau}{T^2}\right)},
\end{equation}
 and is 
\begin{equation} \label{eq58}
  T_{tr}(0.1)=\frac{T^2}{\tau}\log(3).
\end{equation}
Finally, although
Eqs. (\ref{eq57}) were obtained within the weak dephasing approximation, they are valid even in the strong dephasing regime.
In this limit dynamics are completely incoherent and are governed by rate equations \cite{Shore}.
\subsection{General case with $\gamma_{ij}=\gamma$} \label{sec42}
Although analytic solutions for Eqs. (\ref{eq27}) cannot be derived when $\phi(t)$ is no longer constant, 
the main features for the system dynamics can be qualitatively discussed. In order to simplify Eqs. (\ref{eq27}),
we first note that for Gaussian pulses the coupling term $\dot{\phi}\sin\theta$ scales with the relative 
pulse delay $\tau$, see Eqs. (\ref{eq62}) and (\ref{eq63}). 
On the other hand the effective coupling terms $\Omega_{su}(t)$, $\Omega_{sv}(t)$ and $\Omega_{uv}(t)$ 
all scale with the relaxation rate $\gamma$. Then, in the weak dephasing regime these latter coupling terms are negligible
compared to $\dot{\phi}\sin\theta$. Using this approximation the equations for the coherences $u(t)$ and $v(t)$ become
\begin{equation} \label{eq59}
  \begin{split}
    &\dot{u}(t)\approx-\Gamma_u(t)u(t)+2v(t)\dot{\phi}\sin\theta, \\~\\
    &\dot{v}(t)\approx-\Gamma_v(t)v(t)-2u(t)\dot{\phi}\sin\theta.
  \end{split}
\end{equation}
Furthermore, we have that
\begin{equation} \label{eq60}
  \vert\Gamma_s(t)\vert\gg\vert\Omega_{su}(t)\vert,\vert\Omega_{sv}(t)\vert.
\end{equation}
With this the evolution of $s(t)$ and consequently that of the dark states populations,
is dominated by the effective relaxation rate $\Gamma_s(t)$,
\begin{equation} \label{eq61}
  \dot{s}(t)\approx-\Gamma_s(t)s(t).
\end{equation}

Thus, in the weak dephasing regime coherences obey the dynamics of a two-level system Eq. (\ref{eq59}),
while the populations decay exponentially at a rate $\Gamma_s(t)$ (\ref{eq61}). Although analytic solutions 
could be derived for Eq. (\ref{eq59}) following a similar method to that of Sec. \ref{sec41}, the complexity
of the final expressions limits their use. Nevertheless, useful conclusions can be drawn by simple inspection
of these equations. When the relaxation rate $\gamma$ is increasing, and for fixed delays, the coherences
and the populations will decay. Taking into account that $\Gamma_s(\pm\infty)=\Gamma_u(\pm\infty)=0$,
$\Gamma_v(\pm\infty)=\gamma$ and that $\Omega_{ij}(\pm\infty)=0$, we anticipate that in the long time limit, 
$v(t\rightarrow\infty)=0$, whereas both $s(t)$ and $u(t)$ acquire a finite constant value. Thus in the long 
time limit coherences are partially preserved.

The effect for different delay times on the system dynamics is more complicated.
This is due to the dependence that both the coupling $\dot{\phi}\sin(\theta)$ and the effective relaxation rates 
have with respect to the delay time. Taking into account the results from the previous section \ref{sec41} and those
for STIRAP in $\Lambda$-systems \cite{Ivanov2004}, we expect that the dependence of the fidelity with respect to
the pulse delay will reflect the dependence of the transition time $T_{tr}(\epsilon)$ on the pulse delay.  
\section{Numerical simulations} \label{sec5}
We present now results from numerical simulations with the master equation (\ref{eq15}) and for 
Gaussian pulses. For overlapping control and Stokes pulses, 
$\Omega_c(t)=\Omega_s(t)$, we use the pulses given in Eq. (\ref{eq42}). For the pulse ordering where
the Stokes pulse proceeds and ends after the pump pulse, while the control pulse is delayed relative to both pulses, 
the parametrization reads
\begin{equation} \label{eq62}
  \begin{split}
    &\Omega_p(t)=\Omega_0e^{-(t-\tau/2)^2/T^2}, \\ &\Omega_s(t)=\Omega_0e^{-(t-\tau/2)^2/(2T^2)},\\ &\Omega_c(t)=\Omega_0e^{-(t+\tau/2)^2/(2T^2)}.
  \end{split}
\end{equation}
For the pulse ordering Stokes-control-pump we have
\begin{equation} \label{eq63}
  \begin{split}
    &\Omega_p(t)=\Omega_0e^{-(t+\tau/2)^2/T^2},\\ &\Omega_s(t)=\Omega_0e^{-(t-\tau/2)^2/T^2},\\ &\Omega_c(t)=\Omega_0e^{-t^2/T^2}.
  \end{split}
\end{equation}
For the ordering control-Stokes-pump, the pulses are the same as above with the only change that the 
Stokes and control pulses are interchanged, i.e.\ $\Omega_s(t)\rightarrow\Omega_c(t)$ and $\Omega_c(t)\rightarrow\Omega_s(t)$.
\subsection{Overlapping Stokes and control pulses} \label{sec51}
\begin{figure}[tbp]
  \begin{center}
    \resizebox{85mm}{!}{\includegraphics{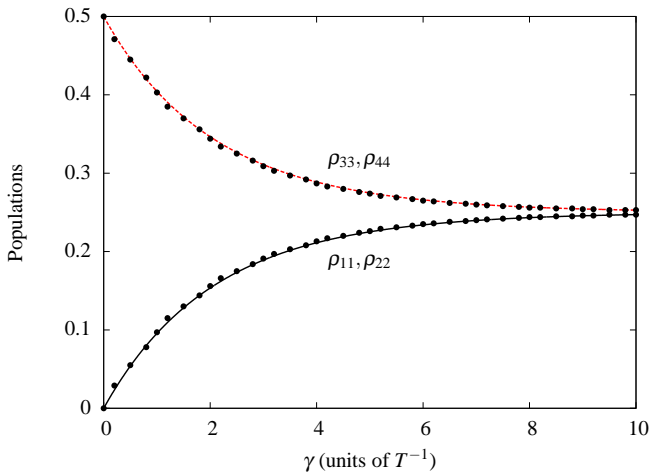}}
    \caption{(Color online) Final populations $\rho_{jj}(\infty)$ for the bare states $\psi_j$ plotted against the
    dephasing rate $\gamma_{ij}=\gamma$ for Gaussian pulses Eq. (\ref{eq42}). The peak Rabi frequency
    is $\Omega_0=50T^{-1}$, and the delay is $\tau=1.5T$. The dots are numeric results and the solid
    lines were obtained from Eqs. (\ref{eq26}) and (\ref{eq57b}), and with the aid of Eqs.  
    (\ref{eq17}), (\ref{eq18}) and (\ref{eq24}).} \label{fig3}
  \end{center}
\end{figure}

Starting with a pulse ordering where the control and Stokes pulse overlap, i.e.\ $\Omega_s(t)=\Omega_c(t)$, in Fig. 
\ref{fig3} we plot the populations $\rho_{jj}(\infty)$ for the bare states $\psi_j$, and for different relaxation rates 
$\gamma_{ij}=\gamma$. The validity of the analytic results of Sec\ref{sec41} is confirmed, where we see that 
the results almost coincide. 
The expected population damping for the dark states due to losses towards 
the bright states, is evidenced as a population drop for the states $\psi_3$ and $\psi_4$. This is accompanied from 
population rise for the other two states $\psi_1$ and $\psi_2$. We also note the predicted asymptotic behavior for strong 
dephasing, where $\rho_{jj}(\infty)\rightarrow1/4$.    

\begin{figure}[tbp]
  \begin{center}
    \resizebox{85mm}{!}{\includegraphics{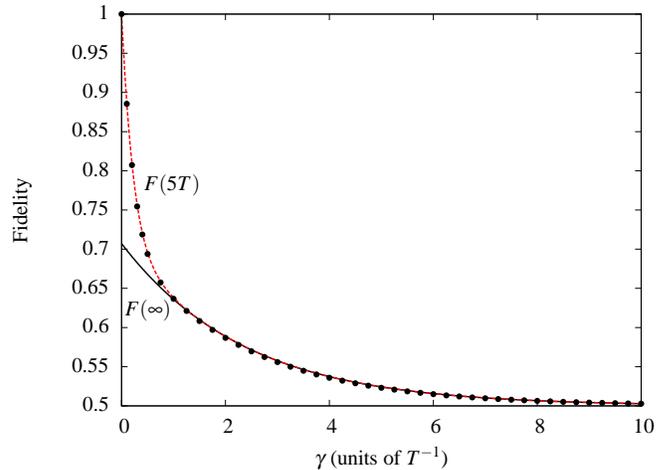}}
    \caption{(Color online) 
      The fidelity for the coherent superposition state (\ref{eq14}) as a function of the dephasing rate $\gamma_{ij}=\gamma$
      for the Gaussian pulses (\ref{eq42}). 
      The peak Rabi frequency is $\Omega_0=50T^{-1}$ and the delay is $\tau=1.5T$. The dots are numeric
      results, whereas the solid line is the analytic result for strong dephasing or for $t\gg1/\gamma$. 
      The dotted line is the fidelity for the weak dephasing regime (\ref{eq64}), and for $t_{max}=5T$.} \label{fig4}
  \end{center}
\end{figure}

Population losses for the dark states will also result in a fidelity decay for the final coherent state Eq. (\ref{eq14}).
For weak dephasing $\gamma T\ll1$ and for times shorter than $1/\gamma$, i.e.\ $T\ll t_{max}\ll 1/\gamma$, 
the fidelity (\ref{eq29}) reads
\begin{equation} \label{eq64}
  F^2(t_{max})=\rho^a_{11}(\infty)+\textrm{Re}\{\rho^a_{12}(t_{max})\},
\end{equation}
where $\rho^a_{11}(\infty)$ is given by Eq. (\ref{eq57b}) and $\textrm{Re}\{\rho^a_{12}(t_{max})\}$ is derived from 
Eq. (\ref{eq57a}), with the substitution
\begin{displaymath}
  \lim_{t\rightarrow\infty}e^{-\gamma t}\rightarrow e^{-\gamma t_{max}}.
\end{displaymath}
In Fig. \ref{fig4}, the fidelity is plotted against the relaxation rate $\gamma_{ij}=\gamma$ for the Gaussian
pulses (\ref{eq42}). The dotted line is the fidelity $F(t_{max})$ for $t_{max}=5T$, where an exponential
drop for increasing $\gamma$ is noted. More specifically, it can be shown that for $\gamma T^2/\tau\ll1$ the fidelity is
\begin{equation} \label{eq65}
  F^2(t_{max})\approx\frac{1}{4}\left(1+e^{-\frac{c_s\gamma T^2}{4\tau}}\right)+\frac{1}{2}e^{-\frac{c_u\gamma T^2}{4\tau}-\gamma t_{max}}.
\end{equation}
On the other hand for strong dephasing $\gamma T^2/\tau\gg1$, or in the long time limit  $t\gg1/\gamma$, the system 
completely decoheres and the only contribution to the fidelity is from the population $\rho^a_{11}(\infty)$ Eq. (\ref{eq57b}), 
i.e.\
\begin{equation} \label{eq66}
  F^2(\infty)=\rho^a_{11}(\infty).
\end{equation}
For this limit the final fidelity is well below unity, see solid line, and it rapidly drops
for increasing dephasing rate.

\begin{figure}[tbp]
  \begin{center}
    \resizebox{85mm}{!}{\includegraphics{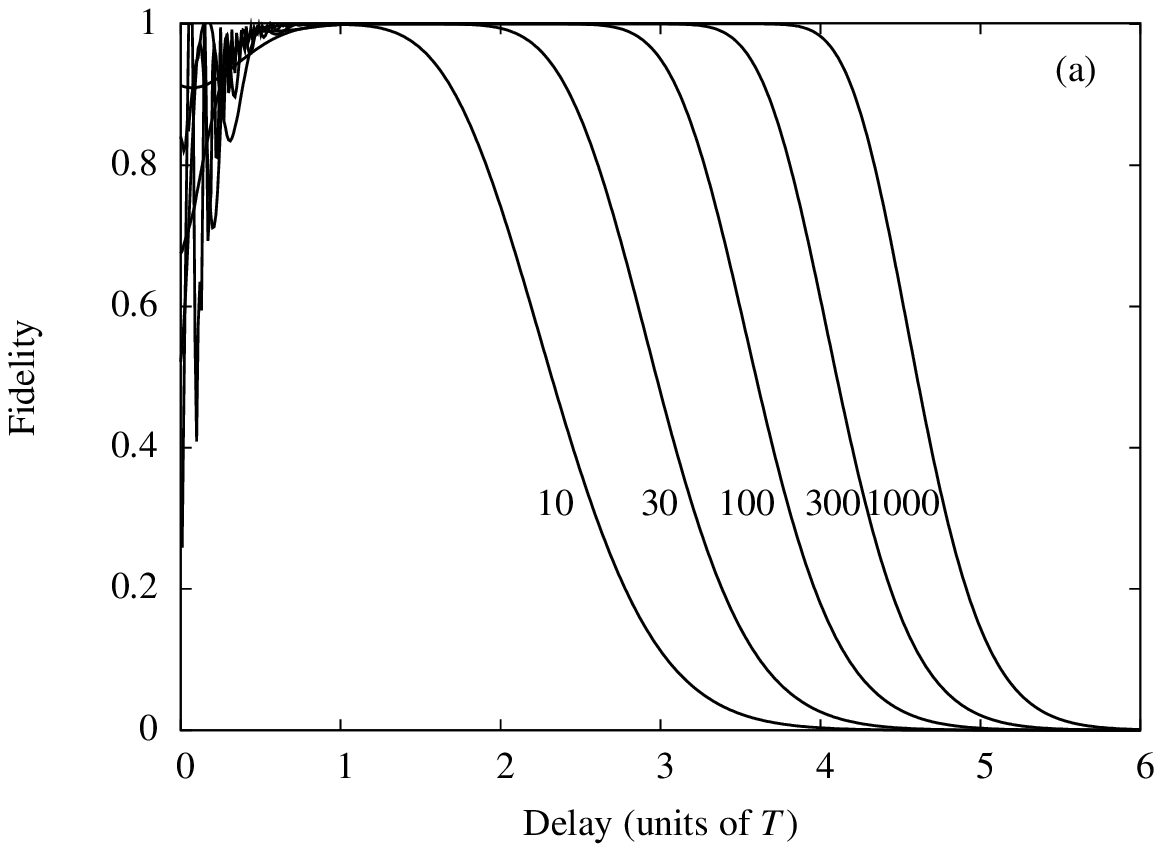}} \\~\\
    \resizebox{85mm}{!}{\includegraphics{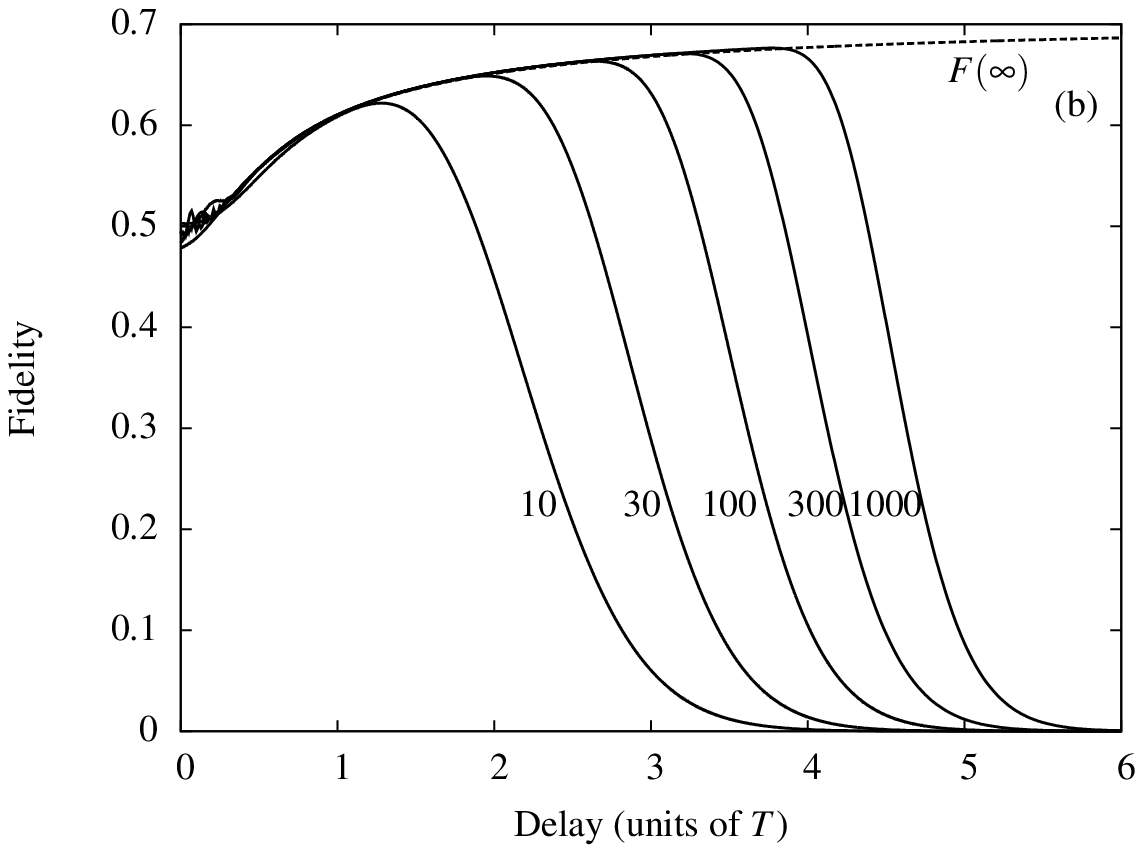}}
    \caption{(a) The fidelity for the coherent state (\ref{eq14}) plotted against the delay time $\tau$ 
      for the Gaussian pulses (\ref{eq42}), as obtained from simulations with the master equation. 
      The peak Rabi frequencies $\Omega_0$ are denoted next to
      the respective curves and $\gamma_{ij}=\gamma=0$. (b) The same with (a) but for $\gamma_{ij}=\gamma=T^{-1}$. The
      dashed line is the analytic result obtained from Eqs. (\ref{eq66}) and (\ref{eq57b}).}\label{fig5}
  \end{center}
\end{figure}

As already noted in Sec. \ref{sec41} the delay $\tau$ is expected to have an inverse effect on the fidelity. This is due to
the inverse dependence of the transition time Eq. (\ref{eq58}) with respect to the delay. This is demonstrated in
Fig. \ref{fig5}. The fidelity for different peak Rabi frequencies is plotted against the delay for $\gamma_{ij}=\gamma=0$,
Fig. \ref{fig5}(a), and for $\gamma_{ij}=\gamma=T^{-1}$, Fig. \ref{fig5}(b). From this we see that for adiabatic evolution 
the fidelity increases as Eqs. (\ref{eq66}) and (\ref{eq57b}) predict (dashed). The regime 
of validity for these equations increases for increasing pulse areas, and this is because of the adiabatic condition.
\subsection{Stokes-control-pump pulse ordering} \label{sec52}
\begin{figure}[tbp]
  \begin{center}
    \resizebox{85mm}{!}{\includegraphics{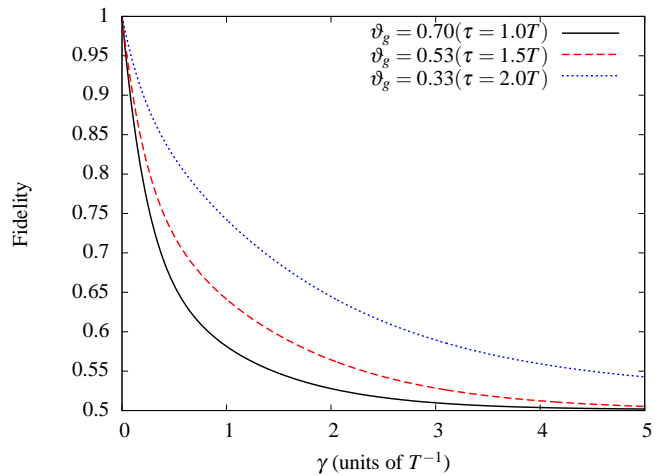}}
    \caption{(Color online) 
      The fidelity for the coherent state (\ref{eq12}) as a function of the dephasing rate $\gamma_{ij}=\gamma$
      for the Gaussian pulses (\ref{eq63}). 
      The peak Rabi frequency is $\Omega_0=200T^{-1}$ and the delay is $\tau=1.0T$ (solid) $\tau=1.5T$ (dashed)
      and $\tau=2.0T$ (dotted). The results were obtained from a numerical simulation with the 
      master equation (\ref{eq15}).} \label{fig6}
  \end{center}
\end{figure}

\begin{figure}[tbp]
  \begin{center}
    \resizebox{85mm}{!}{\includegraphics{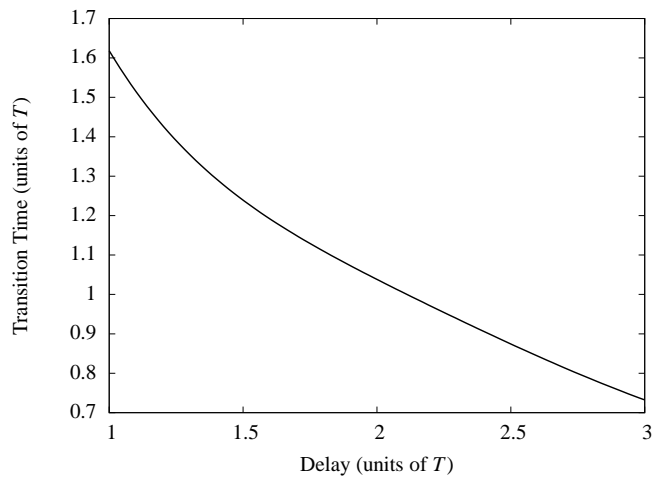}}
    \caption{The transition time $T_{tr}(\epsilon)$ (\ref{eq30}) for the Gaussian pulses (\ref{eq63}). The
      peak Rabi frequency is $\Omega_0=200T^{-1}$ and $\epsilon=0.1$. The times $t_\epsilon$ and $t_{1-\epsilon}$, were derived 
      by numerically solving the equations $F^2(t_{\epsilon})=\epsilon$ and $F^2(t_{1-\epsilon})=1-\epsilon$.} \label{fig7}
  \end{center}
\end{figure}

When considering a pulse ordering where the Stokes proceeds the control and the pump is delayed further 
from the control, Eq. (\ref{eq63}), 
the exact form of the resulting coherent state (\ref{eq12}) will be a function of the delay $\tau$ via the
geometric phase $\vartheta_g$ (\ref{eq7}). This means that the inverse effect that the increasing delay $\tau$
has on the fidelity will reflect differently upon different coherent states. This is shown in Fig. \ref{fig6},
where the fidelity for different coherent states (different $\tau$) as obtained from numerical 
simulations is plotted against the dephasing rate $\gamma_{ij}=\gamma$. It is clear that as we increase the delay 
time and for a given $\gamma$, the fidelity for the corresponding state
\begin{equation} \label{eq67}
  \Psi(\infty)=-\psi_3\sin\vartheta_g(\tau)-\psi_4\cos\vartheta_g(\tau),
\end{equation}
will be higher than the fidelity for a state that corresponds to smaller $\tau$. This, as already said, is because of the
inverse dependence of the transition time (\ref{eq30}) on the delay $\tau$, see Fig. \ref{fig7}. As the relaxation rate
increases, the fidelity for all the coherent states decreases reaching the strong dephasing asymptotic $F=0.5$.
\subsection{Stokes proceeds and ends after the pump pulse} \label{sec53}
\begin{figure}[tbp]
  \begin{center}
    \resizebox{85mm}{!}{\includegraphics{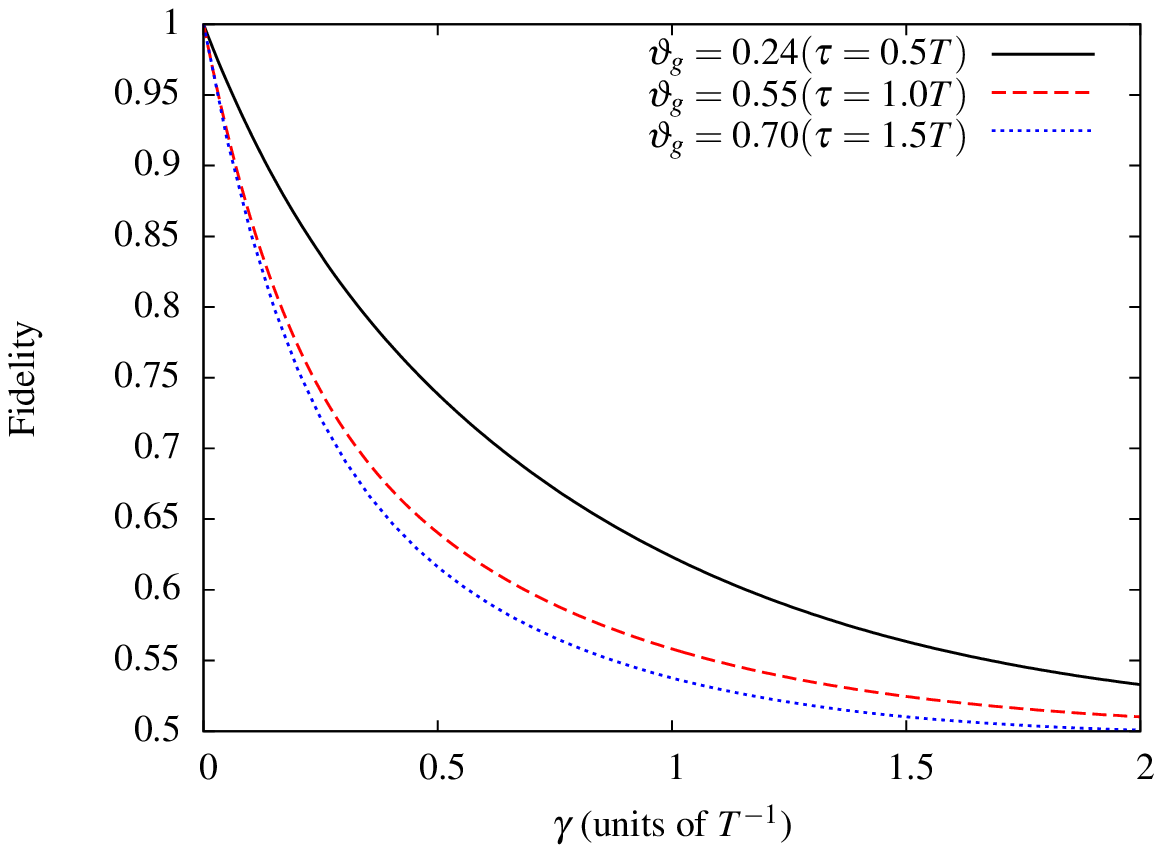}}
    \caption{(Color online) 
      The fidelity for the coherent state (\ref{eq11}) as a function of the dephasing rate $\gamma_{ij}=\gamma$
      for the Gaussian pulses (\ref{eq62}). 
      The peak Rabi frequency is $\Omega_0=200T^{-1}$ and the delay is $\tau=0.5T$ (solid) $\tau=1.0T$ (dashed)
      and $\tau=1.5T$ (dotted). The results were obtained from a numerical simulation with the 
      master equation (\ref{eq15}).} \label{fig8}
  \end{center}
\end{figure}

In contrast to the previous two pulse orderings, this one differs in the effect that the delay $\tau$ has on the
fidelity. Dephasing has the same effect where the fidelity drops exponentially for increasing
$\gamma_{ij}=\gamma$, see Fig. \ref{fig8}. In addition to this increasing the delay will also result in a fidelity 
reduction. The explanation for this is again given in terms of the transition time and its
dependence with respect to the delay.

\begin{figure}[tbp]
  \begin{center}
    \resizebox{85mm}{!}{\includegraphics{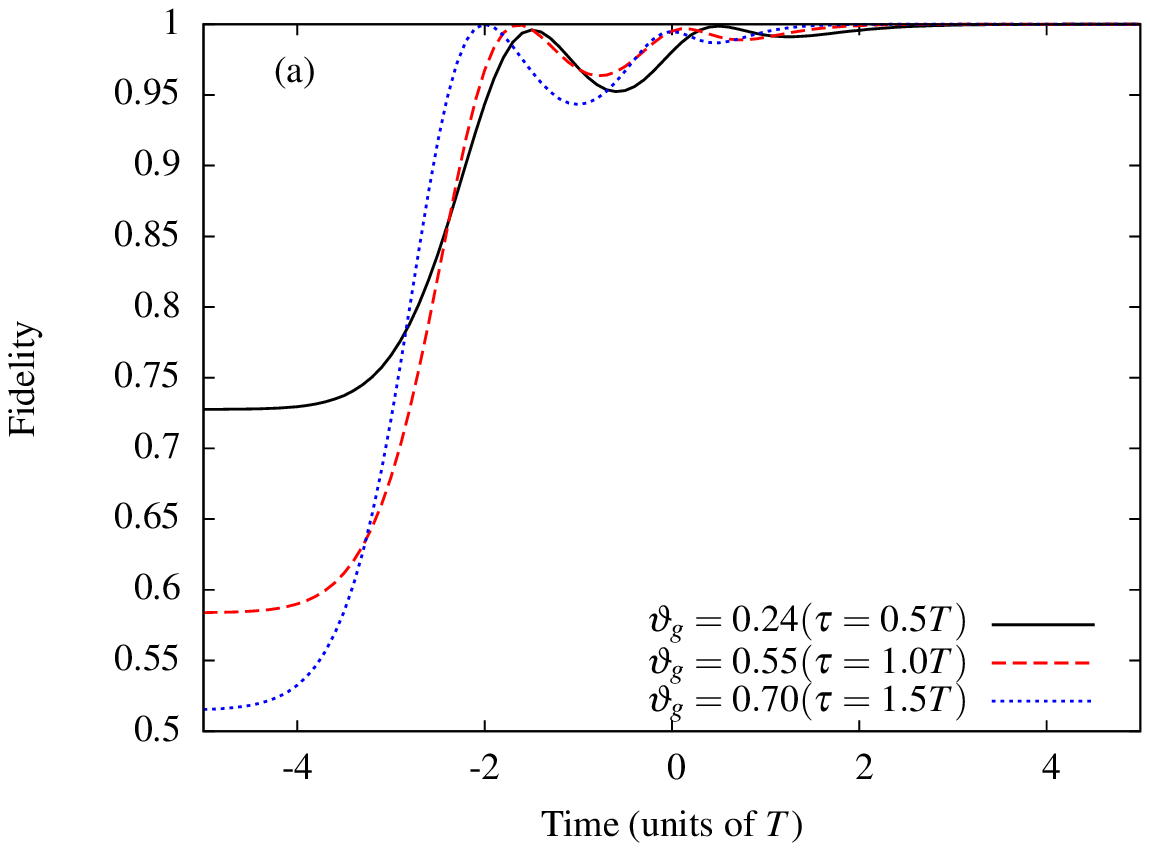}} \\~\\ 
    \resizebox{85mm}{!}{\includegraphics{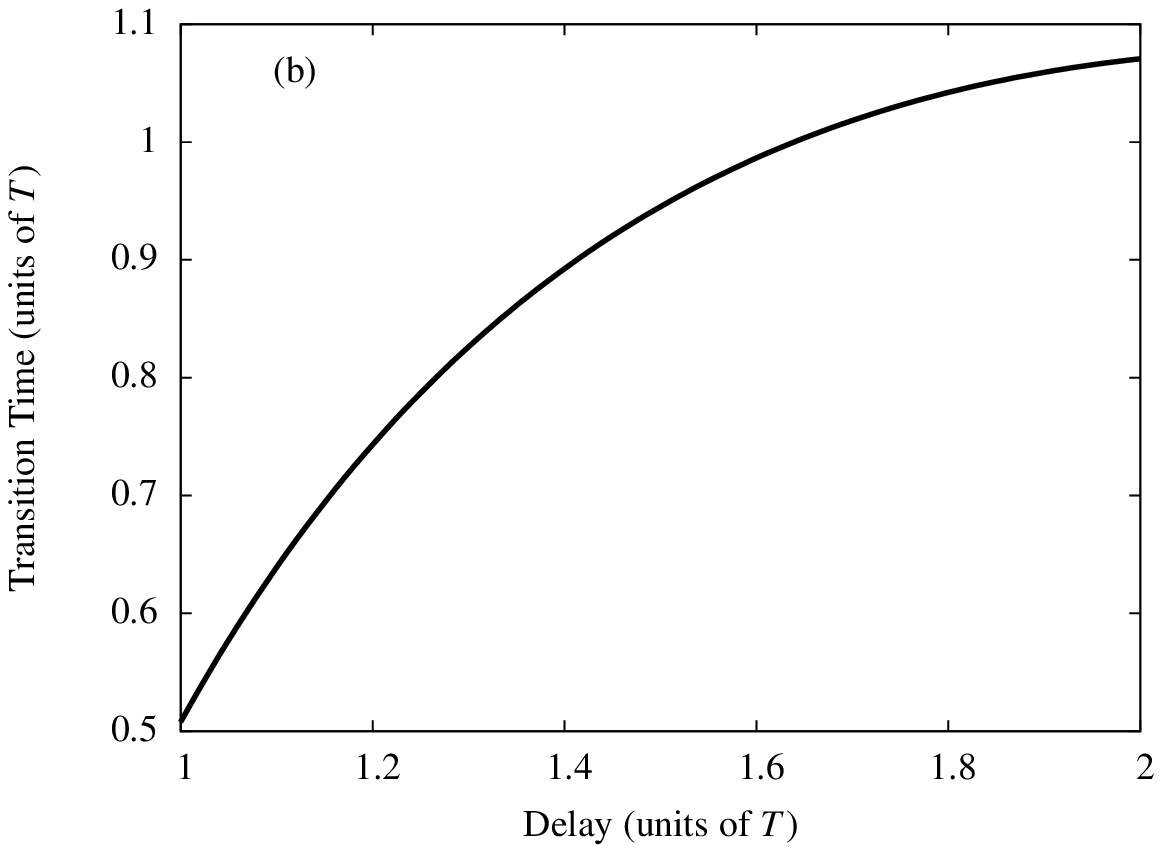}}
    \caption{(Color online) (a) The fidelity for the coherent state (\ref{eq11}) as a function of time 
      for the Gaussian pulses (\ref{eq62}), as obtained from simulations with the master equation. The peak Rabi frequency
      is $\Omega_0=200T^{-1}$, whereas the delay is $\tau=0.5T$ (solid) $\tau=1.0T$ (dashed) and $\tau=1.5T$ (dotted). The 
      relaxation rates are $\gamma_{ij}=\gamma=0$. (b) The transition time $T_{tr}(\epsilon)$ Eq. (\ref{eq31}) for 
      $\epsilon=0.1$. The times $t_{1+\epsilon}$ and $t_{1-\epsilon}$, were derived 
      by numerically solving the equations 
      $F^2(t_{1+\epsilon})=(1+\epsilon)\cos^2(\vartheta_g)$ and $F^2(t_{1-\epsilon})=1-\epsilon$. }\label{fig9}
  \end{center}
\end{figure}

As already pointed out in Sec. \ref{sec33} the fidelity for the coherent state (\ref{eq11}) and for 
$t\rightarrow-\infty$ is $F^2(-\infty)=\cos^2\vartheta_g$. As the geometric phase
increases, i.e.\ the delay increases, the initial value for the fidelity decreases. 
Thus, for increasing $\tau$ the transition time 
is expected to increase too. This is shown in Fig. \ref{fig9}, where the fidelity as a function of time
and for different delays in the absence of decoherence, Fig. \ref{fig9}(a), and the corresponding
transition times, Fig. \ref{fig9}(b), are plotted. As we can see in Fig. \ref{fig9}(a) the initial value for the
fidelity reduces for increasing delays. For this reason the transition
time will increase, Fig. \ref{fig9}(b), and consequently the effect of dephasing on the fidelity increases.
\section{Conclusion} \label{sec6}
In this paper we have studied the effects that dephasing has on STIRAP in tripod configurations.
We have derived an exact adiabatic solution for the master equation in the case of 
overlapping Stokes and control pulses, and for weak dephasing. The results were verified with numerical simulations 
to provide a very accurate approximation for the fidelity dynamics. In the adiabatic limit, population losses and 
dephasing for the dark states depend only on the relative relaxation rates of the three bare ground states, 
but not on those for the intermediate state. 

The fidelity exponentially decreases for increasing dephasing, whereas the pulse delay has an inverse effect. This is
due to the fact that the transition time decreases for increasing delay. This way dephasing effects are suppressed. 
Using numerical simulations we extended our studies to different pulse orderings. For a pulse ordering of the form
Stokes-control-pump, or when interchanging the order of the control and Stokes pulses, similar dynamics are observed 
as with overlapping Stokes and control pulses.
The fidelity decrease exponentially with the dephasing, whereas the delay has again an inverse effect. This will reflect 
differently on each coherent state. This is because the geometric phase, which characterizes the final superposition,
is a function of the delay.

For a pulse ordering where the Stokes pulse proceeds and ends after the pump pulse, while the control pulse follows
the delay has a different effect. The reason for this is that the transition time increases with the delay.
Dephasing has again the same effect leading to an exponential decrease for the fidelity.    
\acknowledgments
This work has been supported by the European Commission's ITN project FASTQUAST and the Bulgarian
NSF grant No D002-90/08.
\appendix
\section{Deriving the equations for the two-level system} \label{appA}
From Eqs. (\ref{eq16}), (\ref{eq17}), (\ref{eq18}), (\ref{eq19}), and (\ref{eq24}) we can derive the following 
equation for the population inversion $w_{34}=\rho^a_{33}-\rho^a_{44}$  
\begin{subequations}\label{eqA1}
  \begin{align}
    &\dot{w}_{34}(t)=\left(\gamma_{12}\sin^2(\theta)+\cos^2(\theta)\Gamma_2(\phi)\right)w_{34}(t), \label{eqA1a}
    \intertext{where}
    &\Gamma_j(\phi)=\cos^2(\phi)\gamma_{j3}+\sin^2(\phi)\gamma_{j4}. \label{eqA2b}
  \end{align}
\end{subequations}
Taking into account the initial condition $\rho^a_{33}(-\infty)=\rho^a_{44}(-\infty)=0$, or $w_{34}(-\infty)=0$, 
we see that that the inversion is $w_{34}(t)=0$, i.e.\ $\rho^a_{33}=\rho^a_{44}$. Making next the substitution
\begin{equation} \label{eqA2}
  \rho^a_{33}=\rho^a_{44}=(1-\rho^a_{11}-\rho^a_{22})/2,
\end{equation}
we get the following master equation for the effective two-level system spanned by the two dark states (\ref{eq2})
\begin{subequations} \label{eqA3}
  \begin{align}
    &\dot{\rho}^a_{kl}=-\mathcal{D}^{ij}_{kl}\rho^a_{ij}-\imu\dot{\phi}\sin(\theta)[\sigma_z,\rho^d]_{kl}-D^0_{kl}, \label{eqA3a}
    \intertext{where $\rho^d$ is}
    &\rho^d=\left(\begin{array}{cc}
      \rho^a_{11} & \rho^a_{12} \\
      \rho^a_{21} & \rho^a_{22} \end{array}\right),
  \end{align}
\end{subequations}
and $\sigma_z$ is the relevant Pauli matrix.
In this equation the Einstein summation convention is used, where $i,j,k,l=1,2$. The tensor $\mathcal{D}^{ij}_{kl}$ 
and the matrix $D^0_{ij}$ resulted when using the rotation matrix $R(t)$ (\ref{eq18}) to transform the 
master equation (\ref{eq15}) in the adiabatic basis Eq. (\ref{eq19}).
At this point their exact analytic form is not important. Instead some very useful properties that are going to be
used next are listed below

\begin{subequations} \label{eqA4}
  \begin{align}
    &(\mathcal{D}^{ij}_{12})^\ast=\mathcal{D}^{ji}_{12}, \label{eqA4a} \\
    &\mathcal{D}^{11}_{21}=\mathcal{D}^{22}_{21},\quad\mathcal{D}^{11}_{12}=\mathcal{D}^{22}_{12}, \label{eqA4b}\\
    &\mathcal{D}^{11}_{11}=\mathcal{D}^{22}_{22},\quad\mathcal{D}^{22}_{11}=\mathcal{D}^{11}_{22}, \label{eqA4c}\\
    &\mathcal{D}^{12}_{11}=\mathcal{D}^{12}_{22},\quad\mathcal{D}^{21}_{11}=\mathcal{D}^{21}_{22}, \label{eqA4d}\\
    &\mathcal{D}^{12}_{12}=(\mathcal{D}^{12}_{12})^\ast=\mathcal{D}^{21}_{21}=(\mathcal{D}^{21}_{21})^\ast,\label{eqA4e}\\
    &\textrm{Im}\{\mathcal{D}^{12}_{11}\}=-\frac{1}{2}\textrm{Im}\{\mathcal{D}^{11}_{12}\}, \label{eqA4f}\\
    &\textrm{Re}\{\mathcal{D}^{12}_{11}\}=\frac{1}{2}\textrm{Re}\{\mathcal{D}^{11}_{12}\},\label{eqA4g} \\
    &D^0_{11}=D^0_{22}=-\frac{1}{4}(\mathcal{D}^{11}_{11}+\mathcal{D}^{22}_{11}), \label{eqA4h} \\
    &D^0_{12}=(D^0_{21})^\ast=-\frac{1}{2}\mathcal{D}^{11}_{12}. \label{eqA4i}
  \end{align}
\end{subequations}

Having these properties we can now proceed and derive the equations for the population inversion $w^a_{12}(t)=\rho^a_{11}-\rho^a_{22}$,
the populations $\rho^a_{11}$ and $\rho^a_{22}$, and the coherences $u(t)=\sqrt{2}\textrm{Re}\{\rho^a_{12}\}$ and 
$v(t)=\sqrt{2}\textrm{Im}\{\rho^a_{12}\}$. Starting from the inversion $w^a_{12}(t)$, is easy to show that
\begin{equation} \label{eqA5}
  \dot{w}^a_{12}(t)=(\mathcal{D}^{22}_{11}-\mathcal{D}^{11}_{11})w^a_{12}(t).
\end{equation}
Taking into account the fact that $\theta(-\infty)=0$ and $\rho_{11}(-\infty)=1$, 
we have that $\rho^a_{11}(-\infty)=\rho^a_{22}(-\infty)=1/2$ and $w^a_{12}(-\infty)=0$. From
the above equation we have that at all times
\begin{equation} \label{eqA6}
  w^a_{12}(t)=0
\end{equation}
and that
\begin{equation} \label{eqA7}
  \rho^a_{11}=\rho^a_{22}=\frac{1}{4}-\frac{s(t)}{2}.
\end{equation}
With this result the remaining three equations for $s(t)$, $u(t)$ and $v(t)$ (\ref{eq27}) are derived,
with the initial conditions being $s(-\infty)=-1/2$, $u(-\infty)=1/\sqrt{2}$ and $v(-\infty)=0$.
The effective relaxation rates and Rabi frequencies read

\begin{subequations} \label{eqA8}
  \begin{align}
    \nonumber \Gamma_s(t)&=\mathcal{D}^{11}_{11}+\mathcal{D}^{22}_{11} \\
    &=\frac{1}{2}\sin^2(2\theta)\Gamma_1(\phi)
    +\frac{1}{2}\cos^4(\theta)\sin^2(2\phi)\gamma_{34}, \label{eqA8a} \\
    \nonumber \Gamma_{u}(t)&=\mathcal{D}^{12}_{12}+\textrm{Re}\{\mathcal{D}^{12}_{21}\} \\
    &=\frac{1}{4}\sin^2(2\theta)\Gamma_1(\phi)
    +\frac{1}{4}\left(1+\sin^2(\theta)\right)^2\sin^2(2\phi)\gamma_{34}, \label{eqA8b} \\
    \nonumber \Gamma_v(t)&=\mathcal{D}^{12}_{12}-\textrm{Re}\{\mathcal{D}^{12}_{21}\} \\
     &=\cos^2(\theta)\Gamma_1(\phi)
    +\sin^2(\theta)\cos^2(2\phi)\gamma_{34}, \label{eqA8c}
  \end{align}
\end{subequations}
and
\begin{subequations} \label{eqA9}
  \begin{align}
    \nonumber \Omega_{su}(t)=&\textrm{Re}\{\mathcal{D}^{11}_{12}\}=\frac{1}{4}\sin^2(2\theta)\Gamma_1(\phi) \\
    &-\frac{1}{4}\cos^2(\theta)\left(1+\sin^2(\theta)\right)\sin^2(2\phi)\gamma_{34}, \label{eqA9a} \\
    \nonumber \Omega_{sv}(t)=&\textrm{Im}\{\mathcal{D}^{11}_{12}\}
    =-\frac{1}{4}\cos^2(\theta)\sin(\theta)\sin(4\phi)\gamma_{34} \\
    &+\frac{1}{2}\cos^2(\theta)\sin(\theta)\sin(2\phi)(\gamma_{14}-\gamma_{13}), \label{eqA9b} \\
    \nonumber \Omega_{uv}(t)=&\textrm{Im}\{\mathcal{D}^{12}_{21}\}=
    -\frac{1}{16}\left(\sin(3\theta)-7\sin(\theta)\right)\sin(4\phi)\gamma_{34}
    \\&-\frac{1}{2}\cos^2(\theta)\sin(\theta)\sin(2\phi)(\gamma_{14}-\gamma_{13}). \label{eqA9c} 
  \end{align}
\end{subequations}
\section{Adiabatic integrals} \label{appB}
In order to obtain $s(\infty)$ and $u(\infty)$, we first use the inverse rotation $\mathcal{R}^{-1}(\infty)$ 
(\ref{eq36}) on the state $\psi(\infty)$ (\ref{eq48}) to get $c_1(\infty)$ and $c_2(\infty)$
(\ref{eq34}). Using the expressions for $c_{\pm}(\infty)$ and Eq. (\ref{eq34}) we have for $s(\infty)$
\begin{subequations} \label{eqB1}
  \begin{align}
    &s(\infty)\propto\exp(I_s) \label{eqB1a}
    \intertext{where $I_s$ is}
    &I_s=\left(\int_{-\infty}^{0}(\epsilon_+(t)-\Gamma_s(t))dt+\int_{0}^{\infty}(\epsilon_{-}(t)-\Gamma_s(t))dt
    \right) \label{eqB1b}
    \intertext{and for $u(\infty)$}
    &u(\infty)\propto\exp(I_u)=\exp\left(\int_{-\infty}^{\infty}(\epsilon_{+}(t)-\Gamma_s(t))dt\right). \label{eqB1c}
  \end{align}
\end{subequations}
Although the expressions for $\epsilon_{\pm}(t)$ and $\Gamma_s(t)$ are very complicated, they can all
be parametrized in terms of single dimensionless variable $x=4t\tau/T^2$, i.e.\
\begin{subequations} \label{eqB2}
  \begin{align}
    &\epsilon_{\pm}(t)=\gamma g_{\pm}(x), \label{eqB2a} 
    \intertext{and}
    &\Gamma_s(t)=\gamma g_s(x), \label{eqB2b}
    \intertext{where}
    &g_-(x)=\frac{4(e^x-1)}{(1+e^x)(2+e^x)^2(1+\sqrt{1+8/(1+e^x)^2})}, \label{eqB2c} \\
    \nonumber \\ 
    &g_+(x)=\frac{(1-e^{2x})(1+\sqrt{1+8/(1+e^x)^2})}{2(2+e^x)^2}, \label{eqB2d} 
    \intertext{and}
    &g_s(x)=\frac{2(1+2e^x)}{(2+e^x)^2}. \label{eqB2e}
  \end{align}
\end{subequations}

With this the integral for $I_s$ in Eq. (\ref{eqB1a}), takes the form
\begin{equation} \label{eqB3}
  \begin{split}
    I_s=&\frac{\gamma T^2}{4\tau}\int_{-\infty}^{0}(g_+(x)-g_s(x))dx \\~
    \\&+\frac{\gamma T^2}{4\tau}\int_{0}^{\infty}(g_-(x)-g_s(x))dx,
  \end{split}
\end{equation}
where both integrals converge, and can be calculated numerically giving
\begin{equation} \label{eqB4}
  I_s=-\frac{c_s\gamma T^2}{4\tau},
\end{equation}
where $c_s=2.42$. When calculating the integral $I_u$ in Eq. (\ref{eqB1b}), attention must be paid 
for $x\rightarrow\infty$. For this limit the integrand converges to a finite value
\begin{equation} \label{eqB5}
  \lim_{x\rightarrow\infty} (g_+(x)-g_s(x))=-1.
\end{equation}
Breaking the integral into three parts we have
\begin{equation} \label{eqB6}
  \begin{split}
    I_u=&\frac{\gamma T^2}{4\tau}\int_{-\infty}^{0}(g_+(x)-g_s(x))dx
    \\~\\&+\frac{\gamma T^2}{4\tau}\int_{0}^{\infty}(g_+(x)-g_s(x)+1)dx \\~\\&-\frac{\gamma T^2}{4\tau}\int_{0}^{x\rightarrow\infty}dx,
  \end{split}
\end{equation}
where in the integrand for $x>0$ we have added and subtracted its asymptotic value for $x\rightarrow\infty$. The 
first two integrals converge, whereas the third one gives rise to the exponential term $e^{-\gamma t}$ in Eq. (\ref{eq57a}). Thus the 
integral $I_u$ reads
\begin{equation} \label{eqB7}
  I_u=-\frac{c_u\gamma T^2}{4\tau}-\gamma t,
\end{equation}
where $c_u=0.68$. 

%
\end{document}